\documentclass[twocolumn,trackchanges]{aastex701}

\newcommand{\hi}{H\textsc{i}}
\newcommand{\slope}{0.80}
\newcommand{\merr}{0.12}
\newcommand{\bintercept}{0.94}
\newcommand{\berr}{0.19}
\newcommand{\rsquared}{0.67}
\newcommand{\rerr}{0.08}

\newcommand{\htwo}{$\mathrm{H}_2$}
\newcommand{\jrd}[1]{\textcolor{black}{#1}}

\newcommand{\edit}[1]{{#1}}

\received{July 29, 2026}
\revised{August 4, 2026}
\accepted{August 5, 2026}
\submitjournal{ApJ Letters}

\begin{document}

\title{Scylla: Observational Evidence for an Order of Magnitude in Dust Mass Opacity Evolution with ISM Density in the Large Magellanic Cloud}

\author[0000-0003-0588-7360]{Christina W. Lindberg}
\affiliation{Center for Astrophysics $\vert$ Harvard \& Smithsonian, 60 Garden St., Cambridge, MA 02138, USA}
\affiliation{Space Telescope Science Institute, 
3700 San Martin Drive, 
Baltimore, MD 21218, USA}
\altaffiliation{JC Ryan Post-Doctoral Fellow}
\email[show]{christina.lindberg@live.com}  

\author[0000-0001-7959-4902]{Christopher J. R. Clark}
\affil{AURA for the European Space Agency, Space Telescope Science Institute, 3700 San Martin Drive, Baltimore, MD 21218, USA}
\email[hide]{cclark@stsci.edu}  

\author[0000-0002-7743-8129]{Claire E. Murray}
\affil{Space Telescope Science Institute, 
3700 San Martin Drive, 
Baltimore, MD 21218, USA}
\affiliation{The William H. Miller III Department of Physics \& Astronomy, Bloomberg Center for Physics and Astronomy, Johns Hopkins University, 3400 N. Charles Street, Baltimore, MD 21218, USA}
\email[hide]{cmurray1@stsci.edu}  

\author[0000-0001-6326-7069]{Julia Roman-Duval}
\email{duval@stsci.edu}
\affiliation{Space Telescope Science Institute, 3700 San Martin Drive, Baltimore, MD 21218, USA}
\email[hide]{duval@stsci.edu}

\author[0000-0001-6118-2985]{Caroline Bot}
\affiliation{Université de Strasbourg, CNRS, Observatoire astronomique de Strasbourg, UMR 7550, F-67000 Strasbourg, France}
\email{caroline.bot@astro.unistra.fr}

\author[0000-0003-1680-1884]{Yumi Choi}
\affiliation{NSF National Optical-Infrared Astronomy Research Laboratory, 950 North Cherry Avenue, Tucson, AZ 85719, USA}
\email{yumi.choi@noirlab.edu}

\author[0000-0002-2970-7435]{Roger E. Cohen}
\affiliation{Department of Physics and Astronomy, Rutgers the State University of New Jersey, 136 Frelinghuysen Rd., Piscataway, NJ, 08854, USA}
\email[hide]{rc1273@physics.rutgers.edu}  

\author[0000-0002-8937-3844]{Steven R. Goldman}
\affil{Space Telescope Science Institute, 
3700 San Martin Drive, 
Baltimore, MD 21218, USA}
\email[hide]{sgoldman@stsci.edu} 

\author[0000-0001-5340-6774]{Karl D. Gordon}
\affiliation{Space Telescope Science Institute, 3700 San Martin Drive, Baltimore, MD 21218, USA}
\affiliation{Sterrenkundig Observatorium, Universiteit Gent, Krijgslaan 281 S9, B-9000 Gent, Belgium}
\email[hide]{kgordon@stsci.edu}

\author[0000-0001-5538-2614]{Kristen B.\ W.\ McQuinn}
\email{kmcquinn@stsci.edu}
\affiliation{Department of Physics and Astronomy, Rutgers, the State University of New Jersey,  136 Frelinghuysen Road, Piscataway, NJ 08854, USA}
\affiliation{Space Telescope Science Institute, 3700 San Martin Drive, Baltimore, MD 21218, USA}
\email[hide]{kmcquinn@stsci.edu}

\author[0000-0003-1356-1096]{Elizabeth Tarantino}
\affiliation{Space Telescope Science Institute, 3700 San Martin Drive, Baltimore, MD 21218, USA}
\email[hide]{etarantino@stsci.edu} 

\author[0000-0002-7502-0597]{Benjamin F.\ Williams}
\email{benw1@uw.edu}
\affiliation{Department of Astronomy, University of Washington, Box 351580, U.W., Seattle, WA 98195-1580, USA}

\author[0000-0002-9912-6046]{Petia Yanchulova Merica-Jones}
\affiliation{University of Sofia, Faculty of Physics, 5 James Bourchier Blvd., 1164 Sofia, Bulgaria}
\affiliation{Institute of Astronomy and NAO, Bulgarian Academy of Sciences, 72 Tsarigradsko Chaussee Blvd., 1784 Sofia, Bulgaria}
\email[hide]{pyanchulova@astro.bas.bg}  

\author[0000-0002-2250-730X]{Catherine Zucker}
\affil{Center for Astrophysics $\vert$ Harvard \& Smithsonian, 60 Garden St., Cambridge, MA 02138, USA}
\email[hide]{catherine.zucker@cfa.harvard.edu}



\begin{abstract}

The emissivity of dust is known to vary greatly with radiative environment, density, grain chemistry, and geometry. Discrepancies between dust mass surface densities derived from far-infrared (FIR) emission and visible extinction persist across and within galaxies in the local Universe. Here, we use new extinction and emission measurements towards the LMC to show that this discrepancy is driven by the dust mass opacity evolving with the intrinsic density of the ISM, and that the ratio between FIR and optical dust mass opacity varies with gas surface density. These new findings imply that the dust mass opacity in the FIR could increase by nearly an order of magnitude (e.g., $\kappa_{160} = 0.3 - 6\ m^2\ kg^{-2}$) across over an order of magnitude of total hydrogen surface density ($\Sigma_H = 4 - 100 M_{\odot}\ pc^{-2}$), corroborating previous theoretical models for dust mass opacity evolution in the FIR, and providing new implications for emission-based dust mass estimates.

\end{abstract}


\section{Introduction} 

Dust grains are a vital component of galaxies for regulating and processing radiation \edit{\citep{draine2003, tielens2005, galliano2018}}. When dust grains are heated by the interstellar radiation field, they reradiate this energy through a variety of emission mechanisms, including photoluminescence in the optical \edit{\citep{witt2004}} and thermal emission in the infrared (IR) \edit{\citep{draine2007, compiegne2011, jones2017}}. Heated dust grains account for the bulk of the emission of galaxies observed at far-IR (FIR) and sub-millimeter (submm) wavelengths \edit{\citep{viaene2016, bianchi2018}}. This emission can be used to measure the physical mass of dust within galaxies \edit{\citep{hildebrand1983, draine2007, galliano2018}}, providing critical insights into their gas mass, interstellar medium (ISM) \edit{\citep{leroy2011, sandstrom2013}}, and star-forming conditions \edit{\citep{kennicutt2012}}.

The observed FIR emission is fundamentally driven by several factors, including dust temperature, total dust mass, and the intrinsic optical properties of dust grains \edit{\citep{draine2003, galliano2018}}. This last term, known as the dust mass absorption coefficient or dust mass opacity ($\kappa_D$), is defined as the efficiency with which dust grains absorb or emit energy per unit mass as a function of wavelength. $\kappa_D$ is dependent on the physical properties of the grains, including chemical composition, morphology, density, and size distribution \edit{\citep{ossenkopf1994, kohler2015, demyk2017II, Demyk2017, ysard2018}}, and is expected to vary across environments as grains evolve from the diffuse ISM to dense molecular cores \edit{\citep{2003A&A...398..551S, kohler2015, jones2017}}. 

$\kappa_D$ is strongly dependent on wavelength. In the optical regime, $\kappa_V$ ($\kappa_D$ at V-band or $0.55 \ \mu m$) is largely dependent on \jrd{grain size/surface area \citep{ysard2018} and will slowly evolve across ISM environments \citep{2016A&A...588A..43J, 2015A&A...580A.136F}.} By contrast, $\kappa_{\it FIR}$ ($\kappa_D$ in the FIR-submm regime) is predicted to increase substantially as grains evolve via coagulation, mantle accretion, and ice formation in denser ISM environments \citep{ossenkopf1994, 2003A&A...398..551S, li2003, kohler2015, jones2017}. 

This environmental dependency leaves $\kappa_{\it FIR}$ poorly constrained in practice, with modeled values and inferred observations spanning nearly two orders of magnitude \citep[e.g., $\kappa$ at $\lambda = 500\ \mu m$, or $\kappa_{500} = 10^{-1} \mbox{--} 10^{1}\, \mathrm{m}^2 \, \mathrm{kg}^{-1}$; see summary in Figure 1 in][]{clark2019}, causing large uncertainties in the estimated dust mass of galaxies. Additionally, recent observations of M74 and M83, nearby active star-forming galaxies, show an inverse correlation between $\kappa_{\it FIR}$ and gas surface density \citep{clark2019}, sparking further confusion regarding the nature of dust grain emission.

In this work, we show empirical evidence for the evolution of $\kappa_{\it FIR}/\kappa_{V}$ across ISM environments by leveraging new measurements of emission and extinction in the LMC \citep{clark2023, lindberg2025}. In Section \ref{sec:data}, we describe the measurements used to derive dust mass surface densities from emission (FIR) and extinction (visible). In Section \ref{sec:results}, we present our findings. In Section \ref{sec:disc}, we discuss the implications of these findings and compare with previous measurements.

\section{Data}
\label{sec:data}

Most existing measurements of $\kappa_{\it FIR}$ are derived by modeling the spectral energy distribution (SED) of FIR-submm emission in environments where the dust mass can be constrained {\it a priori}, for instance, where the dust-to-gas ratio is known from depletion measurements \citep{gordon2014, jrd2017, clark2019}. The sensitivity of this method depends on the number of SED measurements available, the assumptions made during the grain modeling, and the number of components (e.g., gas phases and densities) present within a resolution element \citep{galliano2011}. In particular, because $\kappa_{\it FIR}$ is typically treated as a constant in SED modeling, this fixed assumption fails to account for how grain emissivity evolves across different ISM environments and can drive large systematic offsets in the inferred dust mass surface density ($\Sigma_D$).

Visible extinction ($A_V$) provides an independent constraint on $\kappa_V$ \edit{\citep{draine2003, draine2014}}. Unlike dust emission, which scales with the Planck function and therefore depends on temperature, extinction is set by the grain cross-section relative to incident starlight, a property of the grain itself rather than its thermal state \edit{\citep{ysard2018}}.

The ratio $\Sigma_{D,\ FIR}/\Sigma_{D,\ A_V}$ to serve as a direct proxy for $\kappa_{\it FIR}/\kappa_{V}$. \edit{Both surface densities describe the same physical dust column along a given sightline, and each is recovered by dividing an observable by an opacity that we hold fixed at a single assumed value for every sightline. Were both assumed opacities correct, the two dust masses would agree. Where the masses disagree, the disagreement cannot be attributed to the dust column itself, since that is common to both measurements and divides out of the ratio; it can only reflect the extent to which the assumed opacities misrepresents the true ones.} 

\edit{The ratio $\Sigma_{D,FIR}/\Sigma_{D,A_V}$ is therefore proportional to the true $\kappa_{FIR}/\kappa_V$ along that sightline, with unity indicating that the assumed opacity ratio is correct. What the ratio does not remove is the dust temperature, which enters the emission-based column through the shape of the modified blackbody and remains the leading systematic.} Other sources of scatter, including line-of-sight temperature mixing \citep{shetty2009}, excess submm emission \citep{clark2023}, and grain-size distribution variations, can introduce biases and noise in the $\Sigma_{D,\ FIR}/\Sigma_{D,\ A_V}$ ratio. However, these sources are subdominant to the nearly order-of-magnitude offsets that can occur when assuming a fixed $\kappa_D$ for FIR-submm SED fitting \citep[e.g., $\kappa_{160} = 0.5-4\ m^2\ kg^{-1}$;][]{clark2019}

\subsection{Emission}

To measure $\Sigma_{D,\ FIR}$, we use recently published FIR observations of the LMC \edit{\citep{clark2021, clark2023}}. These observations were constructed by combining high-resolution {\it Herschel} Space Observatory data from the HERITAGE Herschel key project \citep{meixner2013} in five bands ($100\ \mu m$ to $500\ \mu m$), with lower-resolution data from all-sky surveys like Planck \citep{planck2011}, the Infrared Astronomical Satellite \citep[IRAS;][]{neugebauer1984}, and the Cosmic Background Explorer \citep[COBE;][]{boggess1992, silverberg1993}, which capture extended dust emission. By combining these data in Fourier space, these maps capture extended emission that was previously missing from the {\it Herschel} data, resulting in better flux agreement with other telescopes. All the emission maps used in this paper are available on Zenodo at \dataset[https://doi.org/10.5281/zenodo.7392275]{https://doi.org/10.5281/zenodo.7392275}. The feathered Herschel maps can
also be accessed at the NASA/IPAC Infrared Science Archive: \dataset[https://www.ipac.caltech.edu/doi/irsa/10.26131/IRSA545]{https://www.ipac.caltech.edu/doi/irsa/10.26131/IRSA545}.

In the optically thin regime, the specific intensity at frequency $\nu$ is related to the dust mass surface density ($\Sigma_D$) and dust mass opacity ($\kappa_{\nu}$) via a modified blackbody:

\begin{equation}
    I_\nu = \kappa_\nu \, B_\nu(T_d) \, \Sigma_D, 
\end{equation}
\label{eq:fir_emission}

\noindent
where $B_\nu(T_d)$ is the Planck function at the dust temperature $T_d$. To recover $\Sigma_D$, this relation must be jointly fit across multiple FIR-submm bands along with the dust temperature and the wavelength dependence of $\kappa_{\nu}$.

The wavelength dependence of $\kappa_{\nu}$ is typically parameterized by a single emissivity spectral index $\beta$. However, galaxies like the SMC and LMC have been shown to contain excess emission in the submm region \citep{galliano2011, gordon2014}. To account for this excess emission, \citet{clark2023} followed \citet{gordon2014} and used a broken-emissivity modified blackbody (BEMBB) framework with a flexible break wavelength ($\lambda_{\mathrm{break}}$) and two emissivity spectral indices ($\beta$), which modulate $\kappa_D$. At shorter wavelengths, $\kappa_D$ was calibrated using COBE-FIRAS observations of the Milky Way cirrus, such that $\kappa_D(\lambda_{ref}) = 1.24\ m^2\ kg^{-1}$ at $\lambda_{ref} = 160\ \mu m$ \citep{jrd2017}. The wavelength-dependent $\kappa_D(\lambda)$ is then propagated according to the two emissivity spectral indices ($\beta_1$ and $\beta_2$), which ajoin continuously at the break wavelength.
SED models were constructed on a grid with a range of $\Sigma_D$, temperatures ($T_d$), a broken emissivity spectral index ($\beta_{1/2}$), and a corresponding wavelength for the break ($\lambda_{\mathrm{break}}$). In addition to reporting the median and standard deviation of $\Sigma_D$, \citet{clark2023} also provide the full posterior probability distribution for all SED parameters (including $\Sigma_D$), consisting of 1000 posterior samples for each pixel in the convolved emission maps. We use these posteriors to model the distribution of $\Sigma_D$ for each pixel, which we leverage later to bootstrap the correlations between emission- and extinction-based dust mass surface densities and hydrogen surface densities.

The pixel size of the original FIR emission maps in the LMC (e.g., $100\ \mu m$, $250\ \mu m$, $500\ \mu m$) ranged from $3.2^{\prime\prime}$ (arcsec) to $14^{\prime\prime}$. To measure $\Sigma_D$, these maps were convolved and re-projected to $36^{\prime\prime}$, the FWHM resolution at $500\ \mu m$, which corresponds to a spatial resolution of 8.7 parsec (pc) in the LMC. To subtract foreground contamination in emission, \citet{clark2021} used all-sky HI maps from the HI4PI survey \citep{hi4pi2016} to approximate the amount of Milky Way (MW) atomic gas ($N_{HI_{MW}}$) at $0.^{\circ}7$ scales and convert from $N_{HI_{MW}}$ to FIR-submm emission using a polynomial map of dust-to-gas ratios \citep[see Section 6 in][]{clark2021}.

\subsection{Extinction}\label{sec:extinction}

To measure dust mass surface density via extinction, we use new maps towards the LMC derived from the Scylla and METAL surveys \citep{lindberg2026}. Scylla and METAL were Hubble Space Telescope ($HST$) programs that used Wide Field Camera 3 (WFC3) to obtain multi-band imaging of resolved stars across 45 individual fields within the LMC \citep{murray2024b} and additional fields in the SMC. 

Using Bayesian SED modeling \citep{gordon2016}, \citet{lindberg2025} measured line-of-sight extinction towards 1.5 million individual stars using the Scylla and METAL datasets. By isolating extinguished background stars within each field, \citet{lindberg2026} used {\it kriging}, an advanced geostatistical method, to map the total column density of visual extinction ($A_V$) within each field. Extinction maps were sampled at $4^{\prime\prime}$ resolution (0.97 pc). All the extinction maps used in this paper can be found in MAST: \dataset[10.17909/mk54-kg51]{http://dx.doi.org/10.17909/mk54-kg51}.

To propagate the distribution of $A_V$ measurements in each pixel, we assume $A_V$ follows a log-normal distribution, with $\mu^*$ and $\sigma^*$ values for the Gaussian distribution in log-space defined as follows:

\begin{equation}
    \sigma^* = \sqrt{\ln(\sigma^2/\mu^2+1)},
\end{equation}

\begin{equation}
    \mu^* = \ln(\mu) - (\sigma^*)^2/2,
\end{equation}

\noindent
whereby $\mu$ and $\sigma$ are the reported median and standard deviation of each $4^{\prime\prime}$ pixel from the $A_V$ maps, respectively. We sample each pixel 100 times and take the exponent of these values to obtain an $A_V$ distribution for each pixel in each field.

To convert from $A_V$ to $\Sigma_D$, we apply the following relation:

\begin{equation}
A_V = 1.086 \, \kappa_V \, \Sigma_D
\label{eq:av}
\end{equation}

\noindent where $\kappa_V = 3262\ m^2\ kg^{-1}$, as originally assumed in \citet{draine2014}. This $\kappa_V$ corresponds to an average grain size distribution of $\sim 0.1\ \mu m$ based on models from \citet{ysard2018}. \edit{The implications of grain size and metallicity are discussed further in Section \ref{sec:disc}.}


To account for foreground extinction, \citet{lindberg2025} used observations of \hi\ emission at $21\rm\,cm$ from the Galactic All Sky Survey \citep[GASS, which covers the same sky area as HI4PI;][]{mccluregriffiths2009} to identify MW gas towards the LMC, which they converted to $A_V$ assuming a conversion factor calibrated for the diffuse, high-latitude ISM \citep{liszt2014} and $R_V=3.1$. MW foreground extinction towards the observed fields ranges from $A_V = 0.24\mbox{--}0.28$ mag, similar to previous measurements \citep{Subramaniam2010, chen2022}. 

\subsection{Hydrogen Surface Density}

We quantify ISM environments by hydrogen surface density ($\Sigma_H$), the mass of hydrogen present per unit area in units of $M_{\odot}/pc^2$. The $\Sigma_H$ maps were originally constructed in \citet{clark2023} using a combination of atomic \citep[\hi;][]{2003ApJS..148..473K} and molecular gas \citep[CO;][]{2011ApJS..197...16W} data across the entire LMC, isolated from the Milky Way using velocity cuts. While there are certainly caveats to approximating total hydrogen density with proxy tracers like CO (especially at low metallicity), these maps nonetheless provide one of the best ways to quantify how gas surface density varies over several orders of magnitude within the galaxy, and should not considerably alter the subsequent relations derived. 

Based on the original atomic and molecular maps, we assume an RMS uncertainty of $8 \times 10^{19}\ \mathrm{cm^{-2}}$ or $0.64\ M_{\odot}\ pc^{-2}$ for \hi\, and $1.2\mathrm{\ K\ km\ s^{-1}}$ or $7.5\  M_{\odot}\ pc^{-2}$ for \htwo, converted from CO. We resample each pixel in the \hi\ and \htwo\ maps 100 times, assuming a Gaussian distribution with the respective uncertainties above, and combine the two maps to construct 100 separate instances of the total hydrogen surface density maps, $\Sigma_H$.

We convolve the $\Sigma_H$ maps to the limiting data resolution, which corresponds to a FWHM resolution of $60^{\prime\prime}$ (17 pc). To compare with $\Sigma_H$, we convolve and re-project $\Sigma_{D,\,FIR}$ and $\Sigma_{D,\,A_V}$ to the same resolution.

\subsection{Quality Cuts}\label{sec:qc}

To ensure that $\Sigma_{D,\,FIR}$ and $\Sigma_{D,\,A_V}$ probe a well-defined ISM regime, we restrict our analysis to resolution elements with $A_V = 0.6 - 2.0$ mag and remove any resolution elements that were constructed with pixels exceeding $A_V > 3$ mag. These bounds are motivated by the following considerations.

First, MW foreground extinction introduces a systematic uncertainty that becomes comparable to the signal at low $A_V$. While the median MW foreground extinction toward the LMC corresponds to $A_V\approx 0.24-0.28$ mag (Section \ref{sec:extinction}), the dust-to-\hi\ ratio can vary by up to a factor of 2 across the high-latitude diffuse ISM towards the LMC \citep[$4.1 \times 10^{-23} - 8.2 \times 10^{-23}\ M_{\odot}\ pc^{-2}\ cm^2$;][]{clark2023}. Since foreground \hi\ is used to estimate foreground $A_V$, requiring $A_V > 0.6$ mag ensures that the LMC contribution dominates over any foreground uncertainty.

Second, at very low $A_V$, the dust column along a sightline will converge to sample a heterogeneous mixture of low-density ISM, and the relationship between surface density and volume density will break down. Requiring $A_V > 0.6$ mag preferentially selects sight lines where the column is dominated by coherent denser ISM structure, making $\Sigma_D$ a meaningful tracer of the underlying volume density.

Lastly, the upper bound of $A_V < 2.0$ mag reflects the 95th percentile of extinction, above which the distribution becomes increasingly poorly sampled and stochastic \citep{lindberg2026}. We exclude these high-extinction sightlines to ensure that they do not statistically bias our results. We also remove any resolution elements constructed with pixels exceeding $A_V > 3$ mag, the completeness limit of the original extinction maps \citep{lindberg2026}.

\begin{figure}
    \centering
    \includegraphics[width=\linewidth]{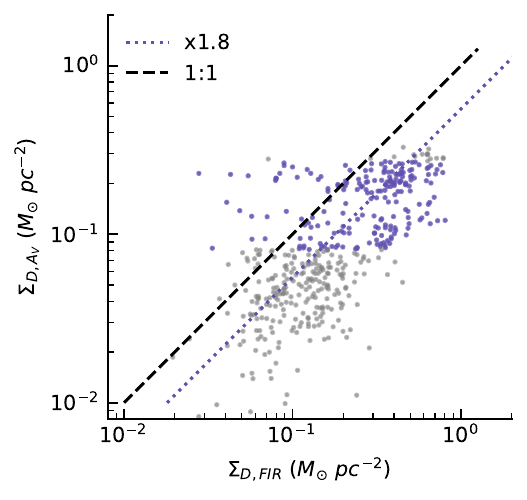}
    \caption{Comparison of median dust mass surface densities derived from FIR emission ($\Sigma_{D,\,FIR}$) and extinction ($\Sigma_{D,\,A_V}$) for all fields in the LMC. Each point represents one sightline with a resolution of $60^{\prime\prime}$/15-pc ($n=510$). Sightlines above ($A_V > 2.0$ mag) or below ($A_V < 0.6$ mag) the imposed quality cuts outlined in Section \ref{sec:qc} are colored grey ($n=296$) and are omitted from subsequent analysis. For visual guidance, we plot a 1:1 trend line (black dashed) and a 1:1.8 trend line (dotted), which depicts the average observed offset between extinction and emission in other galaxies.  }
    \label{fig:dust_mass_ratio}
\end{figure}

\section{Results}
\label{sec:results}

With an accurate $\kappa_D$, $\Sigma_{D,\ FIR}$ should agree with $\Sigma_{D,\ A_V}$. However, we find systematic disagreement across a wide range of surface densities.
In Figure \ref{fig:dust_mass_ratio}, we plot $\Sigma_{D,\ A_V}$ ($\kappa_{V}=3262\ m^2\ kg^{-1}$) against $\Sigma_{D,\ FIR}$ ($\kappa_{160} = 1.24 \, m^2 \, kg^{-1}$) for all fields in the LMC. Each point corresponds to one matched resolution element in each field at $60^{\prime\prime}$ (15 pc). Both quantities span nearly two orders of magnitude, and individual resolution elements show considerable offsets from the 1:1 line (black dashed), with $\Sigma_{D,\ FIR}$ systematically exceeding $\Sigma_{D,\ A_V}$.

The discrepancy between emission and extinction has been noted previously. Comparable factors of 1.8-2.5 offsets, with $\Sigma_{D,\ FIR}$ exceeding $\Sigma_{D, \, A_V}$, have been reported in M31 \citep{dalcanton2015}, in the Milky Way \citep{planck2016}, and in the SMC \citep{ymj2021}. We observe a similar general offset here (blue dotted line). This offset is thought to be driven by variations in dust mass opacity and radiation-field hardness \citep{dalcanton2015}.

\begin{figure}
    \centering
    \includegraphics[width=\linewidth]{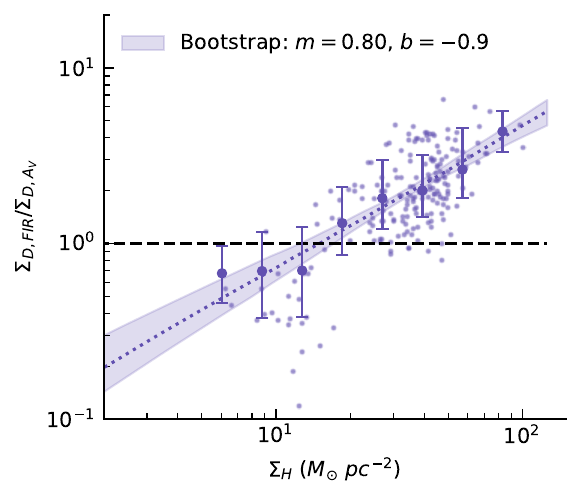}
    \caption{\textbf{Ratio of $\Sigma_D$ versus $\Sigma_H$}: Ratio of extinction and emission-based dust mass surface densities ($\Sigma_{D,\,FIR}/\Sigma_{D,\,A_V}$) versus total hydrogen surface density ($\Sigma_H$), for 208 sightlines with $A_V = 0.6-2.0$ mag, convolved to a resolution of $60^{\prime\prime}$ (15 pc). The y-axis is proportional to $\kappa_{\it FIR}/\kappa_V$.}
    \label{fig:dust_mass_ratio_h}
\end{figure}

While $\Sigma_D$ does not technically inform $\kappa_D$, the ratio of $\Sigma_{D,\,FIR}/\Sigma_{D,\,A_V}$ provides a direct estimate for $\kappa_{\it FIR}/\kappa_V$. 
In Figure \ref{fig:dust_mass_ratio_h}, we compare the ratio $\Sigma_{D,\,FIR}/\Sigma_{D,\,A_V}$ with independent $\Sigma_{H}$. A ratio of one (black dashed line) indicates where the two measurements would be in agreement. Values greater than one suggest that $\Sigma_{D,\,FIR}$ is greater than $\Sigma_{D,\,A_V}$, while values less than one suggest the inverse.  

Observed fields span a factor of 20 in hydrogen surface densities ($\Sigma_H = 4 \mbox{--} 100 \, M_{\odot} \, pc^{-2}$). Shockingly, $\Sigma_{D,\,A_V} / \Sigma_{D,\,FIR}$ spans nearly an order of magnitude, and is moderately correlated with $\Sigma_H$ ($R^2 = \rsquared \pm \rerr$). We perform a simple linear regression fit to characterize the relation between the $\Sigma_D$ ratio and $\Sigma_H$ in the LMC. We repeat this fit 100 times with different samples of $\Sigma_{D,\,A_V}$, $\Sigma_{D,\,FIR}$, and $\Sigma_{H}$, and obtain the following relation for the LMC at $60^{\prime\prime}$ (15-pc) resolution: 

\begin{equation}
\log_{10}\left(\frac{\Sigma_{D,\,FIR}}{{\Sigma_{D,\,A_V}}}\right) = \slope (\pm \merr)\,\log_{10}(\Sigma_H) - \bintercept\ (\pm \berr)
\label{eq:correction}
\end{equation}

\section{Discussion}
\label{sec:disc}

\jrd{Results in Figure \ref{fig:dust_mass_ratio_h} imply that, in high-density environments ($\Sigma_H > 20\ M_{\odot}\ pc^{-2}$), the inferred dust mass surface density from emission ($\Sigma_{D,\,FIR}$) is greater than from extinction ($\Sigma_{D,\,A_V}$). Conversely, in low-density environments ($\Sigma_H < 10 \ M_{\odot}\ pc^{-2}$), $\Sigma_{D,\,FIR}$ is less than $\Sigma_{D,\,A_V}$. Since $\kappa_{V}$ and $\kappa_{\it FIR}$ were both assumed to be constant when fitting $\Sigma_{D,\,A_V}$ and $\Sigma_{D,\,FIR}$, respectively, variations in both contribute to the observed relation across gas densities. }

\jrd{The relative contributions of $\kappa_{V}$ and $\kappa_{\it FIR}$ to the observed trend can be partly disentangled by considering the distinct physical drivers of each. Variations in $\kappa_{\it FIR}$ arise from a broad combination of grain properties, including chemical composition \citep{Demyk2017, demyk2017II} and grain structure such as porosity and aggregation \citep{ysard2018}. By contrast, variations in $\kappa_{V}$ are set primarily by the grain size distribution \citep[Figure 3 in][]{ysard2018}, with $\kappa_{V}$ peaking near a characteristic large-grain size of $\sim 0.1\ \mu m$. Both observations of the LMC extinction curve \citep{2003ApJ...588..871C} and recent simulations \citep{2026MNRAS.tmp..952C} find a median large-grain size of $\sim 0.1\ \mu m$, corresponding to $\kappa_V \approx 3000 \, m^2 \, kg^{-1}$ \citep{ysard2018}, consistent with the value assumed from \citet{draine2014} and applied throughout this work ($\kappa_V = 3262 \, m^2 \, kg^{-1}$; Section \ref{sec:extinction}). Because $\kappa_V$ is maximized near $\sim 0.1\ \mu m$, any increase or decrease in size should result in a lower $\kappa_V$. Across the $A_V = 0.6-2.0$ mag range probed, and at the $\sim 0.5\ Z_{\odot}$ metallicity of the LMC, we anticipate the characteristic large-grain size to grow from $\sim 0.1-0.2\ \mu m$ along diffuse sightlines to $\sim 0.5\ \mu m$ in the densest sightlines. Such growth would reduce $\kappa_V$ by a factor of 3-4 \citep[to $\kappa_V \approx 1000 \, m^2 \, kg^{-1}$;][]{ysard2018} in the densest environments probed. }

\jrd{Since $\Sigma_{D,\,A_V}$ is computed using a fixed $\kappa_V$ (Equation \ref{eq:av}), this assumption directly impacts the ratio in Figure \ref{fig:dust_mass_ratio_h}: in dense sightlines, where the true $\kappa_V$ is lower than assumed, dividing the measured $A_V$ by a too-large $\kappa_V$ results in an under-estimation of $\Sigma_{D,\,A_V}$. Since $\Sigma_{D,\,A_V}$ is in the denominator, this underestimation inflates the ratio precisely in the high-density regime, mimicking a portion of the observed positive trend without any change in $\kappa_{\it FIR}$. However, even when varying $\kappa_V$ from $3000$ to $1000 \, m^2 \, kg^{-1}$ across the hydrogen-density regime probed, we find that grain-size-driven variation of $\kappa_V$ reduced the observed slope only about about half ($m\approx 0.56\pm0.11$ rather than $m = \slope$), indicating that an increase in $\kappa_{\it FIR}$ with density is still required to account for the full relation observed.}

\edit{The analysis above, and our conversion from dust extinction to dust mass (Eq.~\ref{eq:av}), implicitly assume grain composition at Galactic values. The LMC has a metallicity of around 50\% solar and depletes metals differently from the Milky Way, locking proportionally less oxygen and more iron into grains; the silicon and carbon fractions, however, show no significant difference between the two galaxies \citep{jrd2022}. The silicate-to-carbon balance relevant to $\kappa_V$ may therefore differ rather little between them. Even if there were to differ appreciably, the effect on $\kappa_V$ would be limited. Intrinsic opacities vary between grain materials by a factor of a few in the V-band \citep[$3000 \, m^2 \, kg^{-1}$ for a-Sil at $0.1\ \mu m$ versus $10^4 \, m^2 \, kg^{-1}$ for a-C at $0.1\ \mu m$;][]{ysard2018}, but the $\kappa_V$ we observe is a convolution of these opacities with the silicate-to-carbon ratio of the grain population. Consistent with this, the average LMC extinction curve has been observed to be similar to the  Milky Way \citep{gordon2003}. Composition is therefore a secondary influence on $\kappa_V$ relative to grain growth, which varies $\kappa_V$ linearly from $ \approx 1000-3000 \, m^2 \, kg^{-1}$ in the density regime we probe \citep{ysard2018}.}

\subsection{Disagreement with Previous Observations of Dust Mass Opacity Evolution with Gas Densities}
\label{sec:disagreement}

In contrast with our results, \citet{clark2019} used {\it Herschel} data to analyze how $\kappa_D$ varies with ISM environment and found an inverse correlation between $\kappa_D$ and ISM density in the nearby galaxies, M74 and M83 ($D = 5 - 10\ \mathrm{Mpc}$; $\Sigma_{H}=10-40\ M_{\odot}\ pc^{-2}$ for M74 and $\Sigma_{H}=3-600\ M_{\odot}\ pc^{-2}$ for M83). Similarly, \citet{bianchi2019} analyzed the emissivity of nine nearby galaxies ($D = 2.3 - 7.0\ \mathrm{Mpc}$) and also found indications that the emissivity is reduced for galaxies whose gas mass is dominated by \htwo. 

One proposed explanation for this anticorrelation was presented in \citet{priestly2020}, who used a toy model to show that a multi-component dust SED fit could resolve discrepancies in \citet{clark2019} caused by the blending of multiple dust temperatures present along a line of sight. Additionally, \citet{bianchi2022} noted that the discrepancies found at high densities in \citet{bianchi2019} could be the fault of missing CO-dark molecular gas.

So then why are $\kappa_{\it FIR}$ and ISM environment ($\Sigma_H$) not also anticorrelated in the LMC? One potential explanation is the hundred-fold improvement in spatial resolution we achieve in this work (15-pc) compared with M74 and M83 ($\sim1.5$-kpc). At kpc-scales, most ISM structures within galaxies are unresolved\footnote{Previous work from \citet[][see their Fig. 6]{galliano2011} shows that measurements of $\Sigma_{D,\,FIR}$ generally converge at spatial resolutions $< 20$ pc.}, leading to a mixture of dust temperatures, dust masses, and radiative fields observed in an individual FIR-submm emission SED. Subsequently, when SEDs are fit with a single-component dust model, this can introduce degeneracies, leading to an incorrect inference of the $\kappa_{\it FIR}$.
At 15-pc resolution, we have a higher chance of resolving colder ISM structures (e.g., molecular clouds, cold neutral medium) from the warmer ISM, as compared to at 1.5-kpc resolution, leading to an improved inference of $\kappa_{\it FIR}$.  

A further advantage of our approach is that the ratio $\Sigma_{D,FIR}/\Sigma_{D,A_V}$ probes $\kappa_{\it FIR}/\kappa_V$ directly, anchored to a stable optical reference. Studies that infer $\kappa_{\it FIR}$ from emission alone lack this anchor and are therefore more vulnerable to systematic degeneracies between $\kappa_{\it FIR}$, $T_d$, and $\Sigma_D$ in single-component SED fits.

\begin{figure}
    \centering
    \includegraphics[width=\linewidth]{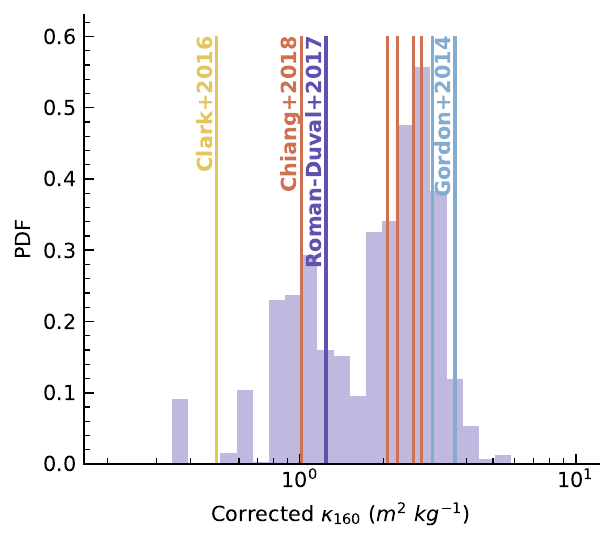}
    \caption{\textbf{Corrected dust mass opacity in the LMC}: Histogram of extinction-corrected dust mass opacity ($\kappa_{160}$) values for all ISM regions measured in the LMC. Compared to existing literature measurements of $\kappa_{160}$, both in the LMC and SMC \citep[e.g.,][]{gordon2014, gordon2017, jrd2017} and beyond \citep[e.g., M101, 22 other galaxies,][respectively]{chiang2018, clark2016}, these new inferred measurements span broader values ($\kappa_{160} = 0.3-6\ m^2 kg^{-1}$). This discrepancy can be explained by the fact that the observed fields probe a wide range of ISM environments ($\Sigma_H = 4 \mbox{--}100\ M_{\odot}\ pc^{-2}$), especially compared to previous surveys that are often biased by measuring $\kappa_D$ towards brighter, higher column density regions of galaxies.}
    \label{fig:kappa_160}
\end{figure}

\subsection{Comparison with Other Dust Mass Opacity Measurements}
\label{sec:kappaD}

The inverse linear relationship between $\kappa_{\it FIR}$ and $\Sigma_{FIR}$ (see Eq.~\ref{eq:fir_emission}) means that we can apply the relative over/underpredictions of $\Sigma_{D,\,FIR}$ to $\kappa_D$\footnote{For example, assuming $\kappa_V$ is correct, if $\Sigma_{D,\,FIR}$ is overpredicted by a factor of two, relative to $\Sigma_{D,\,A_v}$, then $\kappa_D$ must be doubled to compensate for the decrease in dust mass.}, giving us a rudimentary estimate of the correct $\kappa_{\it FIR}$ values present in the LMC. 

All $\Sigma_{D,\,FIR}$ values from \citet{clark2023} were calculated using $\kappa_D(\lambda_{ref}) = 1.24\ m^2\ kg^{-1}$ at $\lambda_{ref} = 160\ \mu m$ ($\kappa_{160}$), based on calibrations from \citet{jrd2017}. Using the $A_V$-calibrated trend observed in Figure \ref{fig:dust_mass_ratio_h} (Eq.~\ref{eq:correction}), we define a corrective factor to $\Sigma_{D,\,FIR}$ and $\kappa_{160}$ as follows:

\begin{equation}
\chi_{160}= 10^{ \slope\ \log_{10}\Sigma_H-\bintercept}
\end{equation} 

\noindent
and apply this correction to $\kappa_{160}$ across all observed $\Sigma_H$ environments. We chose to use $\lambda = {160}\ \mu m$ as the point of comparison, because it was the original calibration wavelength. Additionally, emission at ${160}\ \mu m$ is relatively invariant with wavelength and temperature \citep[as opposed to longer FIR-submm wavelengths,][]{ysard2015, Demyk2017}.

In Figure \ref{fig:kappa_160}, we plot the distribution of corrected $\kappa_{160}$ values within the LMC, all of which had previously been set to $1.24\ m^2\ kg^{-1}$ based on calibrations from \citet[][purple line]{jrd2017}. The locations of fields in the LMC from Scylla target both star-forming regions (e.g., 30$\ $Dor) and more diffuse outskirts of the disk, probing a wide range of ISM environments ($\Sigma_H = 4 \mbox{--}100\ M_{\odot}\ pc^{-2}$).  

We plot previous measurements of $\kappa_{160}$ from the literature using vertical lines in Figure \ref{fig:kappa_160} \citep{gordon2014, clark2016, jrd2017, chiang2018}. The corrected distribution of $\kappa_{160}$ values presented in this work is in broad agreement with previous measurements\footnote{Not all literature values were originally at a reference wavelength of $160\ \mu m$. Those that were not were converted according to Eq.~2 in \citet{clark2019}, assuming $\beta=2$.} and also extends to lower values. 
The one exception to this regional bias comes from \citet{clark2016}, who measured $\kappa_D$ globally for 22 nearby galaxies. The measured $\kappa_{D}$ from \citet{clark2016} ($\kappa_{500} = 0.051\ m^2\ kg^{-1}$) had been considered anomalously low compared to other measurements of $\kappa_{\it FIR}$, until now.
It is possible that by modeling the global SEDs of galaxies, \citet{clark2016} were weighted towards the lower-density ISM, which dominates the total mass, where $\kappa_D$ should be lower.

In future work, we plan to leverage similar extinction measurements to perform a density-dependent recalibration of $\kappa_D$ to refit the observed dust SEDs from {\it Herschel} in the LMC, and expand this work to other nearby galaxies with panchromatic photometry and {\it Herschel} observations, like M31 and M33 \citep{2012ApJS..200...18D, 2021ApJS..253...53W}. This would also recalibrate estimates of how the dust-to-gas ratio varies across ISM environments, as discussed in \citet{clark2023}. 

\section{Conclusions}

In this work, we present empirical evidence for the evolution of the dust mass opacity ($\kappa_D$), as a function of ISM density in the LMC. By comparing dust mass surface densities derived independently from FIR emission ($\Sigma_{D,\,FIR}$) and optical extinction ($\Sigma_{D,\,A_V}$), we find a moderate correlation between the ratio of these two measurements and the total hydrogen surface density ($\Sigma_{H}$). This ratio serves as an indirect proxy for $\kappa_{\it FIR}/ \kappa_{V}$.

\begin{enumerate}
    \item \textit{Emission v.~Extinction} - We find systematic offsets between $\Sigma_{D,\,FIR}$ and $\Sigma_{D,\,A_V}$. Discrepancies match previously observed $1.8\mbox{--}2.5$ factor offsets (e.g., Milky Way, SMC, M31), whereby $\Sigma_{D,\,FIR}$ exceeds $\Sigma_{D,\,A_V}$. (\S\ref{sec:results}; Figure \ref{fig:dust_mass_ratio})  

    \item \textit{Dust Mass Ratio v.~Hydrogen} - We find a moderate correlation between $\Sigma_{D,\,FIR}/\Sigma_{D,\,A_V}$ and $\Sigma_H$, suggesting dust is more emissive per unit mass in high column density ISM environments ($\Sigma_H > 20\ M_{\odot}\ pc^{-2}$) and less emissive in low column density environments ($\Sigma_H < 10\ M_{\odot}\ pc^{-2}$). These findings directly support theoretical predictions that $\kappa_{\it FIR}$ increases in denser regions, tracing dust grain evolution. (\S\ref{sec:results}; Figure \ref{fig:dust_mass_ratio_h}, Equation \ref{eq:correction})  

    \item \textit{Disagreement with Previous Literature} - The positive correlation between $\kappa_D$ and gas densities presented in this work runs counter to recent observations of other galaxies like M74 and M83, which found negative correlations. We attribute this inversion to the hundred-fold improvement in spatial resolution achieved in this study (15-pc vs 1.5-kpc), which better resolves ISM structures and leads to improved temperature inference when fitting emission. (\S\ref{sec:disagreement})  

    \item \textit{Corrections to $\kappa_{\it FIR}$} - Applying a corrective factor derived from the $\Sigma_{D}$ ratio-vs-$\Sigma_{H}$ relation, we generated a distribution of extinction-corrected $\kappa_{160}$ values for the LMC ($\kappa_{160} = 0.3 - 6\ m^2\ kg^{-1}$). This distribution is in broad agreement with $\kappa_{160}$ measurements inferred from other galaxies ($\kappa_{160} = 0.5-4\ m^2\ kg^{-1}$), but spans an even broader distribution.
    (\S\ref{sec:kappaD}, Figure \ref{fig:kappa_160})  
    
\end{enumerate}

\begin{acknowledgments}

Support for this work was also provided through program HST GO-15891, provided by NASA through a grant from the Space Telescope Science Institute, which is operated by the Association of Universities for Research in Astronomy, Inc., under NASA contract NAS 5-26555. The work of Y. Choi is supported by NSF NOIRLab, which is managed by AURA under a cooperative agreement with the U.S. National Science Foundation.  

CWL thanks Kim-Vy Tran for hosting the CfA ECR Writing Workshop that prompted the initial draft of this paper, and Courtney Watson for providing peer-review feedback on early versions of the draft. 

We acknowledge the use of Claude and Claude Code (Anthropic, model Claude Opus 4.7 and earlier models) in suggesting textual improvements to the narrative flow of this manuscript. 

This work made use of the following software packages: \texttt{astropy} \citep{astropy:2013,astropy:2018,astropy:2022,astropy_21262391}, \texttt{matplotlib} \citep{Hunter:2007}, \texttt{numpy} \citep{numpy}, \texttt{python} \citep{python}, and \texttt{scipy} \citep{2020SciPy-NMeth,scipy_20615351}.

This research has made use of the Astrophysics Data System, funded by NASA under Cooperative Agreement 80NSSC21M00561.

Software citation information aggregated using \texttt{\href{https://www.tomwagg.com/software-citation-station/}{The Software Citation Station}} \citep{software-citation-station-paper,software-citation-station-zenodo}.

\end{acknowledgments}

\facilities{\textit{HST(WFC3)}, {\it Herschel}, {\it Planck}, {\it IRAS}, \textit{COBE}}

\bibliography{sample701}{}
\bibliographystyle{aasjournalv7}

\appendix

\section{Cross-check with Other Emission-based Dust Map}
\label{app:chastenet}

The result presented in Section~\ref{sec:results} uses the far-infrared (FIR) dust mass surface densities ($\Sigma_{D, FIR}$) of \citet{clark2021}. Since the absolute $\Sigma_{D, FIR}$ depends on the adopted dust model and SED-fitting methodology, we test whether the trend between $\Sigma_{D, FIR}/\Sigma_{D, A_V}$ and the total hydrogen surface density $\Sigma_H$ (Figure~\ref{fig:dust_mass_ratio_h}; Equation~\ref{eq:correction}) persists when we substitute an independent emission-based dust map. We use the LMC dust mass surface density map of \citet{chastenet2019}, which was constructed from a different combination of instruments, a different dust model, and a different fitting framework.

\citet{chastenet2019} modeled the dust emission of the Magellanic Clouds on a $\sim10$~pc scale by fitting the mid-infrared (MIR) through submm SED. Their data combine \textit{Spitzer} photometry from the SAGE survey \citep[IRAC 3.6, 4.5, 5.8, and 8.0~\micron, and MIPS 24 and 70~\micron;][]{meixner2006} with \textit{Herschel} photometry from the HERITAGE key project \citep[PACS 100 and 160~\micron, and SPIRE 250, 350, and 500~\micron;][]{meixner2013}, for a total of eleven photometric bands spanning 3.6--500~\micron. All images were convolved to the SPIRE 500~\micron\ resolution ($\sim36\arcsec$) with the \citet{aniano2011} kernels and projected onto a grid with approximately independent point-spread-function sampling, giving a
pixel size of $42\arcsec$ ($\sim10$~pc at the distance of the LMC). MW foreground emission was removed by converting an \hi\ column-density map to a dust column
and subtracting it, following \citet{gordon2014}.

The SED in each pixel was fit with the physical dust model of \citet{draine2007} (adopting the MW $R_V = 3.1$ grain mixture) using the Bayesian \texttt{DustBFF} tool \citep{gordon2014}. The model has five free parameters: the minimum radiation-field intensity ($U_{\rm min}$), the PAH mass fraction ($q_{\rm PAH}$), the fraction ($\gamma$) of the dust mass heated by a power-law distribution of starlight intensities, the dust mass surface density ($\Sigma_D$), and a stellar surface-brightness scaling ($\Omega_\star$) that accounts for direct stellar light in the shortest-wavelength bands. We use the resulting $\Sigma_D$ map in units of $M_{\odot}\ pc^{-2}$.

The two emission-based dust maps differ in two important respects: wavelength coverage and assumed dust model.

\citet{chastenet2019} include shorter-wavelength
\textit{Spitzer} coverage, which better constrains the dust temperature, but use the original
\textit{Herschel} maps before the diffuse-emission restoration of \citet{clark2021}. The original \textit{Herschel} maps recover less extended emission, especially at the shortest FIR wavelengths ($100\ \mu m$ and $160\ \mu m$), so they are less sensitive to dust in diffuse regions.

For the dust models, \citet{chastenet2019} use the physical \citet{draine2007} grain model, while \citet{clark2023} use a BEMBB model to fit the observed emission. Their assumed opacities differ slightly ($\kappa_D \approx 0.9$ versus $1.24\ m^2\ kg^{-1}$ at $160\ \micron$), which would offset the absolute $\Sigma_D$ normalization and subsequent intercept, but not the slope. More importantly, the BEMBB allows the emissivity index $\beta$ to break, whereas the \citet{draine2007} model does not. This flexibility becomes more relevant at low metallicities, where excess 500~\micron\ emission is often observed.

\begin{figure}
  \centering
  \includegraphics[width=0.5\textwidth]{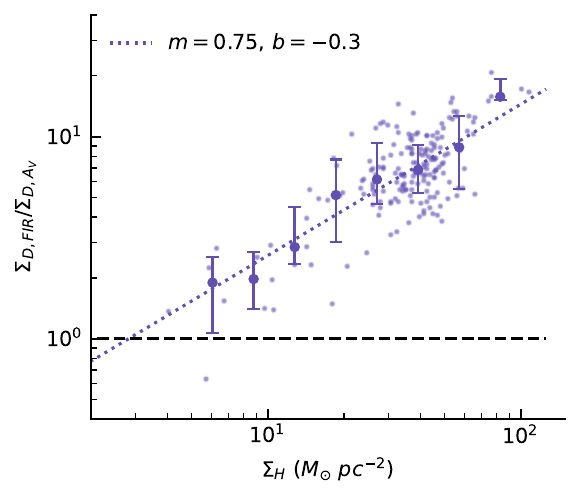}
  \caption{As Figure~\ref{fig:dust_mass_ratio_h}, but using the \citet{chastenet2019} emission-based dust mass surface density map for $\Sigma_{D, FIR}$. The positive trend of $\Sigma_{D, FIR}/\Sigma_{D, A_V}$ with $\Sigma_H$ is recovered with this independent dust emission map.}
  \label{fig:chastenet_ratio}
\end{figure}

We hold the extinction ($\Sigma_{D, A_V}$) and hydrogen ($\Sigma_H$) measurements fixed and substitute the \citet{chastenet2019} $\Sigma_D$ map for $\Sigma_{D, FIR}$, applying the same quality cuts ($A_V = 0.6-2.0$ mag) and the same $60\arcsec$ (15~pc) resolution as in Section~\ref{sec:results}. Because the \citet{chastenet2019} map does not include per-pixel posteriors, we fit a single power and only report the mean relation without an associated uncertainty:

\begin{equation}
\log_{10}\left(\frac{\Sigma_{D, FIR}}{\Sigma_{D, A_V}}\right)
  = 0.75\ \log_{10}(\Sigma_H) - 0.34,
\label{eq:chastenet_fit}
\end{equation}

\noindent with a correlation coefficient $R^2 = 0.53$ (Figure~\ref{fig:chastenet_ratio}).

Despite the different instruments, dust model, and fitting methodology, the \citet{chastenet2019} map recovers a positive slope ($m=0.75$), consistent with our main result ($\slope \pm \merr$; Equation~\ref{eq:correction}), and a comparable correlation strength. Because the two dust models adopt different absolute opacity normalizations, the intercept is offset relative to the \citet{clark2021} analysis. However, the slope, which encodes the density dependence of $\kappa_{FIR}/\kappa_V$, is reproduced independently. This consistency strengthens the conclusion that $\kappa_{\it FIR}/ \kappa_{V}$ increases with ISM density, and indicates that the trend is not an artifact of any single emission-based dust mass estimate.

\end{document}